\documentstyle[12pt]{article}
\date{}
\newcommand{\dmtr}[4]{{
{\left|\begin{array}{cc}#1 & #2\\#3 & #4\end{array}\right|}}}
\def\Bbb#1{{\bf #1}}
\def\eqnum#1{\eqno (#1)}
\title 
{Algorithms for the solution of systems of linear equations in
commutative ring
  \thanks {This paper was  published in the book
  {\em Effective Methods in  Algebraic Geometry}, ed. by
  T. Mora  and C. Traverso,  Progress  in Mathematics 94,
  Birkhauser, 1991, 289--298.
  No part of this materials may be reproduced, stored in retrieval system,
  or transmitted, in any form without prior permission of the copyright owner.
  }
}
\author{Gennadi.I.Malashonok}
\begin{document}
\maketitle
\abstract{
Solution methods for linear equation systems in a  commutative
ring are discussed. Four methods are compared, in  the  setting  of
several different rings: Dodgson's method [1], Bareiss's method [2]
and two methods of the author  -  method  by  forward  and  back-up
procedures [3] and a one-pass method [4].

We show that for the number of coefficient operations, or  for
the number of operations in the finite rings, or for modular computation 
in the polynomial rings the one-pass method [4] is the best.
The method of forward and back-up procedures [3] is  the  best  for
the polynomial rings when we make use of classical  algorithms  for
polynomial operations.}

\section{Introduction}
{\small
Among the set of known algorithms for the solution of  systems
of linear equations, there is a subset, which allows  us  to  carry
out computations within the commutative ring generated by the 
coefficients of the system. Recently, interest in these algorithms grew
due to computer algebra computations. These algorithms may be  used
(a) to find exact solutions of systems with numerical coefficients,
(b) to solve systems over the rings of polynomials with one or many
variables over the integers or over the reals, (c) to solve systems
over finite fields.}

Let
$$
Ax = a  \eqnum{1}
$$
be the given system of linear equations over a commutative ring ${\bf R},$ \
$A\in {\bf R}^{n\times (m-1)}, a\in {\bf R}^{n}, x\in {\bf R}^{m-1}, n<m$,  
with  extended  coefficient  matrix
$A^{*}= (a_{ij})$, $i=1,\ldots,n,$ \  $j=1, \ldots, m$.

The solution of such a system  may  be  written  according  to
Crammer's rule $x_{i}=(\delta ^{n}_{{\rm im}}-\sum ^{m-1}_{j=n+1}
x_{j}\delta ^{n}_{ij}) / \delta ^{n}, $ $ i=1\ldots n,$
where $x_{j}, \ j=n+1\ldots m$,
are free variables and $\delta ^{n}\neq 0.$ \ $ \delta ^{k}$=| $a_{ij}|,$ \  
$i=1\ldots k, \  j=1\ldots k,  \ k=1\ldots n, $ 
denote corner minors of the matrix {\it A} of order $k, \delta ^{k}_{ij}$ 
denotes minors
obtained after a substitution in the minors $\delta ^{k}$ of the column  
$i$  by
the column $j$ of the matrix $A^{*}, \ k=1\ldots n, \ i=1\ldots k, 
\ j=1\ldots  m.$

We examine four algorithms [1]-[4] for solving system (1) over
the fraction field of the ring ${\bf R}$, assuming that ${\bf R}$  does  
not  have
zero divisors and all corner minors $\delta ^{k},k=1\ldots n$, of the matrix
{\it A}  
are
different from zero. Each of the algorithms is in fact a method for
omputing the minors $\delta ^{n}$ and $\delta ^{n}_{ij}$ in the ring  ${\bf R}$.  
For  each  of  the
algorithms we evaluate:

 1.The general time for the solution taking into consideration  only
arithmetic operations and assuming moreover, the execution time for
multiplication, division and addition/subtraction of two  operands,
the first of which is a minor of order {\it i}  and  the  second  one,  a
minor of order $j$, will be $M_{ij}, D_{ij}, A_{ij}$ correspondingly.

 2. The exact number of operations of multiplications, division  and
addition/subtraction over the coefficient of the system.

 3. The number of operations of multiplication/division $(M^{R})$,  
when ${\bf R}={\Bbb R}[x_{1}\ldots x_{r}]$ is a ring of polynomials with $r$  
variables  with  real coefficients and only one computer word is required 
for storing any one of the coefficients.

 4. The number of operations of multiplication/division $(M^{Z})$,  
when ${\bf R}={\Bbb Z}[x_{1}\ldots x_{r}]$ is a ring of polynomials with $r$ 
variables with  integer coefficients and these coefficient are stored in as  
many  computer words as are needed.

 5. The number of operations of multiplication/division $(M^{ZM})$,  
when ${\bf R}={\Bbb Z}[x_{1}\ldots x_{r}]$ is a ring of polynomials with $r$ 
variables with  integer coefficients, but for the solution the modular method  
is  applied, which is based on the remainder theorem.

\eject

\section{ Dodgson's algorithm}

Dodgson's algorithm [1], the first  algorithm  to  be  examined,
appeared more than 100 years before the others.

The first part of the algorithm consists of $n-1$  steps.  In  the
first step, all minors of order 2 are computed
$$
\hat a ^{2}_{ij} = 
a_{i-1,j-1}a_{ij} - a_{i-1,j}a_{i,j-1} ,\ \ \ i=2\ldots n, \ j=2\ldots m,
$$
\noindent 
formed from four neighboring elements, located at the  intersection
of lines $i-1$ and $i$ and of columns $j-1$ and $ j$.

In the $k$-th step, $k=2\ldots n-1,$ according to formula
$$
\hat a^{k+1}_{ij} = 
(\hat a^{k}_{i-1,j-1} \hat a^{k}_{ij} - \hat a^{k}_{i-1,j} \hat a^{k}_{i,j-1})
/ \hat a^{k-1}_{i-1,j-1}, \eqnum{2} 
$$
$$
i=k+1\ldots n, \ j=k+1\ldots m,
$$
\noindent 
all minors  $\hat a^{k+1}_{ij}$ of order $k+1$  are computed, these 
minors are formed
by the elements located at the intersection of the lines 
$ i- k \ldots i$ and
columns $j- k \ldots j$ of the matrix $A^{*}$.

In this way the minors $\hat a^{n}_{nj}=\delta ^{n}_{nj},$ \ 
$j=n\ldots  m$, will be  computed.  Note
that, it is essential that  all  minors  $\hat a^{k}_{ij}, k=1 \ldots n-1,$ 
 $i= k \ldots  n-1$, \ $j= k \ldots m-1$, appearing  in  the  denominator  
of  expression  (2), be different from 0. 
This, of course, narrows down the set of solvable problems.

The second part of Dodgson's algorithm is  feasible  only  under
the assumption that $m=n+1$ and all the searched for unknown variables 
$x_{i}$ belong to ${\bf R}$. Then the value 
$x_n= \delta^n_{nm}/\delta^n_{nn}$  may  be  computed
and substituted in the initial system. After eliminating  the  last
equation and recalculating the column of free members, a system  of
order $n-1$ is obtained, to which the algorithm described above   may
be applied again to find $x_{n-1}$. During this process, it suffices  to
recalculate only those minors in which the column of  free  members
appears. This process continues until all solutions are obtained.

Let us evaluate the computing time of the algorithm. The first
part of the algorithm is executed in time
$$
T^{1}_{1}=(n-1)(m-1)(2M_{1,1}+A_{2,2})+
$$
$$
\sum ^{n-1}_{i=2}(n-i)(m-i)(2M_{i,i}+A_{2i,2i}+D_{2i,i-1}).
\eqnum{3}
$$
The second part of the algorithm, when $m=n+1,$ is  executed  in
time
$$
T^{2}_{1}=\sum ^{n-1}_{i=1}i(M_{1,1}+A_{2,1})+
+\sum ^{n-2}_{j=1}j(2M_{1,2}+A_{3,3})+
$$
$$
\sum^{n-2}_{i=2}\sum^{i}_{j=1}(2M_{i,i+1}+A_{2i+1,2i+1}+D_{2i+1,i-1}).
$$
In this we suppose that the time needed for the  recalculation
of one free member is $M_{1,1}+A_{2,1}$.

\section{Bareiss's Algorithm}

The forward procedure of Bareiss's algorithm [2] differs  from
Dodgson's algorithm only in the selection of the leading (pivoting)
minors.

In the first step all minors of second order are computed
$$
a^{2}_{ij} = a_{11}a_{ij} - a_{1j}a_{i1},\ \  i=2\ldots n, \ j=2\ldots m,
\eqnum{4}
$$
which  surround  the  corner  element $a_{11}$.  At  
the $k$-th   step, $k=2\ldots n-1,$ and according to the formula
$$
a^{k+1}_{ij} = (a^{k}_{kk}a^{k}_{ij} - a^{k}_{ik}a^{k}_{kj})/ a^{k-1}_{k-1,k-1},
\eqnum{5}
$$
$$
i=k+1\ldots n, \ \ j=k+1\ldots m,
$$
\noindent all the minors $a^{k+1}_{ij}$ of order $k+1$ are computed, which are  formed  by
surrounding the corner minor $\delta ^{k}$ by row $ i$ and column $j$, that is, 
minors which are formed by the elements, located at the  intersection
of row $1 \ldots k,i$ and of columns $1 \ldots k, j$.  
Obviously, $a^{n}_{nj}=\delta ^{n}_{nj},$ \  $j= n \ldots  m$, holds.

Comparing the above procedure with Dodgson's algorithm, it  is
seen that here, it suffices that only  the  corner  minors 
$\delta ^{k}=a^{k}_{kk}, \ k=1\ldots n-1,$ be different from zero and 
zero divisors . In order to do so,
they  can  be controlled by choice of the pivot row or column.

The back-up procedure consists of $n-1$ steps, where at the $k$-th
step, $k=1\ldots n-1,$ all the minors

$$
\delta ^{k+1}_{ij} = (a^{k+1}_{k+1,k+1}\delta ^{k}_{i,j} - 
a^{k+1}_{k+1,j}\delta ^{k}_{i,k+1})/ a^{k}_{k,k},
\eqnum{6}
$$
$$
i=k+1\ldots n, j=k+1\ldots m,
$$
are computed.

The computing time of the forward procedure  is  the  same  as
that of the first part of Dodgson's algorithm (3), $T^{1}_{2} = T^{1}_{1}$.

The computing time of the back-up procedure is
$$
T^{2}_{2}=\sum ^{n-1}_{i=1} i(m-i-1)(2M_{i,i+1}+A_{2i+1,2i+1}+D_{2i+1,i}).
$$
\eject
\section{Algorithm of Forward and Back-up Procedures}

The forward procedure in this algorithm [3] is the same as the
forward procedure in Bareiss's algorithm, and is based on  formulae
(4) and (5).

The back-up procedure is more economical, and consists of ${\rm im}$ 
mediate computation of the values $\delta ^{n}_{ij}$ according to the formulae
$$
\delta ^{n}_{ij} = {a^{n}_{nn}a^{i}_{ij} -
\sum ^{n}_{k=i+1} a^{i}_{ik}\delta ^{n}_{kj}\over a^{i}_{ii}}
\eqnum{7}
$$
$$
i=n-1,\ldots 1,\ \ j=n+1 \ldots m.
$$
The computing time of the forward procedure  is  the  same  as
that of the first part of Dodgson's algorithm (3), $T^{1}_{3} = T^{1}_{1}$.

The computing time of the back-up procedure is
$$
T^{2}_{3} =(m-n)\sum ^{n-1}_{i=1} ((n+1-i)M_{n,i}+(n-i)A_{i+n,i+n}+D_{i+n,i}).
$$
Let us note that when $m=n+1$, $T^{2}_{3}$ --- is a quantity of order $n^{2}$, 
and $T^{2}_{2}$ --- a quantity of order $n^{3}$.

\section{One-pass Algorithm}

Algorithms [2] and [3] consist of two parts (two passes). They
make zero elements under  the  main  diagonal  of  the  coefficient
matrix during the first pass. And during the second pass, they make
zero elements up to the main diagonal.

Algorithm [4] requires one pass, consisting of $n-1$  steps.  In
this algorithm, we make diagonalisation of the  coefficient  matrix
minor-by-minor and step-by-step.

In the first step the minors of second order are computed
$$
\delta ^{2}_{2j} = a_{11}a_{2j} - a_{21}a_{1j} , \ \ j=2\ldots m,
$$
$$
\delta ^{2}_{1j} = a_{1j}a_{22} - a_{2j}a_{12} , \ \ j=3\ldots m.
$$
In the $k$-th step, $k=2\ldots n-1,$ the minors of order $k+1$ are compu%
ted according to the formulae (and see the Appendix)
$$
\delta ^{k+1}_{k+1,j} = a_{k+1,k+1}\delta ^{k}_{kk} -
\sum ^{k}_{p=1} a_{k+1,p}\delta ^{k}_{pj}, j=k+1\ldots m,
\eqnum{8}
$$
$$
\delta ^{k+1}_{ij} = (\delta ^{k+1}_{k+1,k+1}\delta ^{k}_{i,j} - 
\delta ^{k+1}_{k+1,j}\delta ^{k}_{i,k+1})/ \delta ^{k}_{k,k},
\eqnum{9}
$$
$$
i=1\ldots k, j=k+2\ldots m.
$$
In this way, at the $k$-th step the coefficients  of  the  first
$k+1$ equations of the system take part.

The general computing time of the solution is
$$
T_{4}= (2m-3)(2M_{1,1}+A_{2,2}) + 
\sum ^{n-1}_{k=2}(m-k) ((k+1)M_{k,1}+ kA_{k+1,k+1}) +
$$
$$
+ \sum ^{n-1}_{k=1}k(m-k-1)(2M_{k,k+1}+A_{2k+1,2k+1}+D_{2k+1,k}).
$$

\section{Evaluation of the Quantity of Operations over the System Coefficients}

We begin the comparison of the algorithms considering the general number of 
multiplication $NM^{m}$, divisions $NM^{d}$ and additions/btractions $NM^{a}$, 
which are necessary for the solution of the  system of linear equations (1) 
of order $n\times m$. Moreover, we  will  not  make
any assumptions regarding the  computational  complexity  of  these
operations; that is we will consider that during the  execution  of
the whole computational process, all multiplications of the coefficients 
are the same, as are the same all divisions  and  all  additions/subtractions.

The  quantity of operations, necessary for Bareiss's algorithm
will be
$$
NM^{m}_{2} = 2n^{2}m-n^{3}-2nm+n, 
$$
$$
NM^{d}_{2} = (2n^{2}m-n^{3}-4nm+2m+3n-2)/2,
$$
$$
NM^{a}_{2} = (2n^{2}m-n^{3}-2nm+n)/2.
$$
The quantity of operations, necessary  for  the  algorithm  of
forward and back-up procedure, is
$$
NM^{m}_{3}=(9n^{2}m-5n^{3}-3nm-3n^{2}-6m+8n)/ 6, 
$$
$$
 NM^{d}_{3}=(3n^{2}m-n^{3}-3nm-6n^{2}+13n-6)/ 6
$$
$$
NM^{a}_{3} =(6n^{2}m-4n^{3}-6nm+3n^{2}+n)/ 6.
$$
The quantity of operations, necessary for the  one-pass  algorithm, is
$$
NM^{m}_{4}=(9n^{2}m-6n^{3}-3nm-6m+6n)/6, 
$$
$$
NM^{d}_{4}=(3n^{2}m-2n^{3}-3nm-6m+2n+12)/ 6
$$
$$
NM^{a}_{4} =(6n^{2}m-4n^{3}-6nm+3n^{2}+n)/ 6.
$$
In the case when the number of equations and unknowns  in  the
system is the same and equal to $n$, we can compare  all  four  algorithms.

\bigskip
\noindent
{\centerline{
\begin{tabular}{|c|c|c|c|}
\hline
\multicolumn{4}{|c|}{quantity of operations}\\
\hline
\# & multiplication & division & add./substr. \\
\hline
1 & $(2n^3 -n^2 - n)\over 2$ & $(n^3 -4n^2 - 5n -2)\over 2$  & $(n^3-n)\over 2$\\
\hline
2 & $ n^3-n$ & $(n^3-2n^2+n)\over 2$ & $(n^3-n)\over 2$ \\
\hline
3 &  $(4n^3 +3n^2 - n-6)\over 6$
  &  $(2n^3 -6n^2 +10n -6)\over 6$ 
  &  $(2n^3+3n^2-5n)\over 6$ \\
\hline
4 & $(n^3 +2n^2 - n-2)\over 2$ & $(n^3 -7n +6)\over 6$ & $(2n^3+3n^2-5n)\over 6$ \\
\hline
\end{tabular}
}}

\bigskip
In this way, according to this evaluation,  the  fourth  algorithm 
(one-pass) is to be preferred. Bareiss's and Dodgson's  algorithms 
are approximately equal regarding  the  quantity  of  operations, 
and each one of them requires three times more divisions and
two times more multiplications, as does the one-pass algorithm. The
third algorithm lies somewhere in between.

If we evaluate according to the general quantity of  multipli%
cation and division operations, considering only the  third  power,
then we obtain the evaluation $3n^{3}/2 : 3n^{3}/2 : n^{3} : 2n^{3}/3.$

\section{Evaluation of the Algorithms in the Ring 
${\Bbb R}[x_{1},x_{2}\ldots x_{r} ]$}

Let ${\bf R}$ be the ring  of  polynomials  of $r$  variables  over  an
integral domain and let us suppose that every  coefficient $a_{ij}$  of
the system (1) is a polynomial of degree $p$ in each variable
$$
a_{ij}= \sum^{p}_{u=0} \sum^{p}_{v=0} \ldots \sum ^{p}_{w=0}
a_{u,v \ldots w}^{ij}x^{u}_{1}x^{v}_{2}\ldots x^{w}_{r} .
\eqnum{10}
$$
Then it is possible to define, how much time is  required  for  the
execution of the arithmetic operations over polynomials  which  are
minors of order $i$ and $j$ of the matrix $A^{*}$
$$
A_{ij}=(jp+1)^{r}a_{ij},
$$
$$
M_{ij}=(ip+1)^{r}(jp+1)^{r}(m_{ij}+a_{i+j,i+j}),
$$
$$
D_{ij}=(ip-jp+1)^{r}(d_{ij}+(jp+1)^{r}(m_{i-j,j}+a_{ii})).
$$
Here we assume, that the classical algorithms for polynomial 
multiplication and division are used. And also,  we  consider  that  the
time necessary for execution of the arithmetic  operations  of  the
coefficients of the polynomials, - when the first operand is  coefficient 
of the polynomial, which is a minor of  order  {\it i},  and  the
second - of order $j$, - is $m_{ij}, d_{ij}, a_{ij}$, for the operations of  
multiplication, division and addition/subtraction, respectively.

Let us evaluate the computing time for each of the four  algorithms, 
considering that the coefficients of  the  polynomials  are
real numbers and each one is stored in one computer word.  We  will
assume that $a_{ij}=0, m_{ij}=d_{ij}=1, A_{ij}=0, 
M_{ij}=i^{r}j^{r}p^{2r}, D_{ij}=(i-j)^{r}j^{r}p^{2r}$,
and we will consider only the leading terms in $m$ and $n$:
$$
M^{R}_{2}= \rho n^{r}\bigg({3m \over 2r+1}- {3n\over 2r+2}\bigg),
$$
$$
M^{R}_{3}= \rho n^r\bigg({3m \over 2r+1}- {3n+3m\over 2r+2}+{3n \over 2r+3}
+ {m-n\over r+1}-{m-n \over r+2}\bigg),
$$
$$
M^{R}_{4}= \rho n^{r} \bigg({3m \over 2r+2}- {3n\over 2r+3}\bigg)
+\rho \bigg({m \over r+2}- {n\over r+3}\bigg),
$$
where $\rho = n^{r+2}p^{2r}$.

For $m=n+1$ it is  possible  to  compare  all  four  algorithms:
$N^{R}_{1}=3\sigma, N^{R}_{2}=(2r+3)\sigma, N^{R}_{3}=2\sigma,
N^{R}_{4}=(2r+1)\sigma +\rho n/(r+2)(r+3)$, where
$\sigma =3n^{2r+3}p^{2r}/(2r+1)(2r+2)(2r+3)$. For $r=0$ we 
obtain the same  evaluation as in the previous section.   
For $r\neq 0$ we obtain $3:(2r+3):2:(2r+1)$.

\section{Evaluation of the Algorithms in the Ring 
${\Bbb Z}[x_{1},x_{2}\ldots x_{r} ]$, Classical Case}

As before we suppose that every coefficient of the system is a
polynomial of the form (10). However,  the  coefficients  of  these
polynomials are now integers and each  one  of  these  coefficients
$a^{ij}_{uv\ldots w}$ is stored in $l$ computer words. 
Then, the coefficients of the
polynomial, which is a minor of order $i$, are integers of length $il$
of computing words.

Under the assumption that classical algorithms  are  used  for
the  arithmetic  operations  on  these  long  integers,  we  obtain:
$a_{ij}=2jla$, $m_{ij}=ijl^{2}(m+2a)$, $d_{ij} =(il-jl+1)(d+jl(m+2a))$, 
where $a, m, d$
are the execution time  of  the  single-precision  operations  of
addition/subtraction, multiplication, and division.

Assuming that $a=0, m=d=1,$ we obtain the  following  evaluation
of the execution times of polynomial  operations: 
$M_{ij}=ijl^{2}(ijp^2)^{r},
D_{ij}=(i-j)^{r+1}j^{r+1}l^{2}p^{2r}, A_{ij}=0.$

In this way, the evaluation of the time for solution  will  be
the same as that for the ring ${\bf R}={\Bbb R}[x_{1}, x_{2}, \ldots, x_{r}]$
(section  6),  if  we
replace everywhere $r$ by $r+1$ and $p^{r}$ by $lp^{r}$.

Therefore, for $m=n+1$  we  obtain $N^{Z}_{1}=3\psi,  N^{Z}_{2}=(2r+5)\psi, 
N^{Z}_{3}=2\psi, N^{Z}_{4}=(2r+3)\psi,$ where 
$\psi =3n^{2r+5}p^{2r}l^{2}/(2r+3)(2r+4)(2r+5),$  
$l\ge 1, r\ge 0.$

\section{Evaluation of the Algorithms in the Ring 
${\Bbb Z}[x_{1},x_{2}\ldots x_{r} ]$, Modular Case}

Let us evaluate the time for the solution of the same problem,
for the ring of polynomials with $r$ variables with integer coefficients 
${\bf R}={\Bbb Z}[x_{1}\ldots x_{r}]$, when the modular method is applied,
based on the
remainder theorem. In this case we will not take into consideration
the operations for transforming the problem in the modular form and
back again.

It suffices to define the number of moduli,  since  the  exact
quantity of operations on the system coefficients for the case of a
finite field has already been obtained in section 4.

We will consider that every prime  modulus $m_{i}$  is  stored  in
exactly one computer word, so that, in order to be able  to  recapture 
the polynomial coefficients, which  are  minors  of  order $n$,
$n(l+\log (np^{3})/2\log m_{i})$  moduli are needed. 
It is easy  to  see  due  to Hadamar's inequality.

Further, we need up moduli for each unknown $x_{j}$, which  appears
with maximal degree $ np$. There are $r$ such unknowns,  and  therefore,
in all, $\mu = prn^{2}(l+\log (np^{3})/2\log m_{i})$ moduli are needed.

If we now make use of the table in section 5, denote the  time
for modular multiplication by $m$ and the time for  modular  division
by $d$, then not  considering  addition/subtraction  and  considering
only leading  terms  in $n$,  we  obtain  for $m=n+1$: 
$N^{ZM}_{1}=(6m+3d)\nu,
N^{ZM}_{2}=(6m+3d)\nu, N^{ZM}_{3}=(4m+2d)\nu, N^{ZM}_{4}=(3m+d)\nu $, 
where $\nu =\mu n^{3}/3.$

\section*{Conclusion}

We see, that modular methods are better then non-modular  ones,
as usual. And the one-pass method [4] is the best for modular  
computation.

The  method of forward and back-up procedures [3]  are  better
for non-modular computation  in  polynomial  rings.  And  Dodgson's
method [1] stands nearly it for such polynomial computations.

\section*{Appendix: Foundation of the One-pass Algorithm}

Identity (8) is an expansion of the minor $\delta ^{k+1}_{k+1,j}$ according  to
line $k+1,$ and therefore it suffices to prove only identity (9).

In order to do so, let us consider the  following  determinant
identity
$$
\dmtr{ A_{si}} {A_{00}}    {N_{s0}} { A_{ij}}=
\dmtr{ A_{0i}} {N_{-t,-j}} {N_{s0}} { A_{ij}}
$$
where $s,i,t,j$ are column numbers of the matrix $A^{*}, A_{uv}$  is  a  
submatrix of the matrix $A^{*}$, this submatrix of order $k$ will be  at  the
left upper corner of a matrix $A^{*}$, if we replace columns $s$ and i by
columns $u$ and $v$ respectively. Here $u=0$ denotes a column  consisting
of zeros, and $u=-t$ denotes a column obtained by changing the  signs
of all the elements of column $t$. Matrix $N_{uv}$ is obtained from matrix
$A_{uv}$ if, an addition, all remaining $k-2$ columns are replaced by zero
ones.

In order to obtain the determinant on the right  side,  it  is
necessary in the determinant on the left side to subtract from  the
first (block) line the second one.

If we expand each one of the determinants of order $2k$  by  the
first $k$ lines according to  Laplace's  rule,  then  we  obtain  the
columns-substitution identity
$$
\delta ^{k}\delta ^{k}_{st;ij} = 
\delta ^{k}_{st}\delta ^{k}_{ij} - \delta ^{k}_{sj}\delta^{k}_{it},
\eqnum{11}
$$
where $\delta ^{k}_{st;ij}$ is a minor formed from the minor $\delta ^{k}$ 
after substituting columns $s$ by $t$ and i by $j$ in the matrix $A^{*}$.

To prove identity (9) it remains to expand the existing minors
of order $k+1$ by row $k+1$
$$
\delta ^{k+1}_{ij} = a_{k+1,k+1}\delta ^{k}_{ij} - 
a_{k+1,j} \delta ^{k}_{i,k+1} - 
\sum ^{k}_{s=1,s\neq i} a_{k+1,s}\delta ^{k}_{s,k+1;i,j}
$$
$$
\delta ^{k+1}_{k+1,j} = a_{k+1,j}\delta ^{k} - 
\sum ^{k}_{s=1} a_{k+1,s}\delta ^{k}_{s,j},
$$
and make use of the columns-substitution identity (11).

\bigskip
\eject
\centerline{REFERENCE}

\bigskip
\noindent 1. Dodgson C.L., Condensation of determinants, being a new and brief
method for computing their arithmetic values,  Proc.  Roy.  Soc.
Lond. A. ( 1866), V.15, 150--155.

\noindent 2. Bareiss E.N., Sylvester's identity and multistep integer-preserving
Gaussian elimination, Math. Comput. ( 1968),  V.22, 565-- 578.

\noindent 3. Malaschonok G.I., On the solution of a linear equation system over
commutative ring,  Math. Notes of  the  Acad.  Sci.  USSR. (1987),
V.42, N4, 543--548.

\noindent 4. Malaschonok G.I., A  new  solution  method  for  linear  equation
systems  over  the  commutative  ring, in  "Int.   Algebraic   Conf.,
Novosibirsk", Aug. 21--26, 1989, Theses on the ring theory,  algebras
and modules, p.82.
\end{document}